%% file: Master.tex
\documentclass[tightenlines,a4paper,12pt,aps]{revtex4}
\usepackage{equations}
\usepackage{epsfig}
\usepackage{graphicx}

\newenvironment{equation*}{\begin{displaymath}}{\end{displaymath}}
\newcommand{\const}{\mbox{const} }
\newcommand{\Scri}{\mbox{$\cal J$}}

\newcommand{\DIII}{\,{}^{\scriptscriptstyle(3)\!\!\!\:}\nabla}
\newcommand{\tDIII}{\,{}^{\scriptscriptstyle(3)\!\!\!\:}\tilde{\nabla}}
\newcommand{\DeIII}{\,{}^{\scriptscriptstyle(3)\!\!\!\:}\Delta}
\newcommand{\RIII}{\,{}^{\scriptscriptstyle(3)\!\!\!\:}R}
\newcommand{\tRIII}{\,{}^{\scriptscriptstyle(3)\!\!\!\:}\tilde{R}}

\newcommand{\ROI}{\,{}^{\scriptscriptstyle(0,1)\!\!\!\:}\hat R}
\newcommand{\RII}{\,{}^{\scriptscriptstyle(1,1)\!\!\!\:}\hat R}
\newcommand{\epsIII}{\,{}^{\scriptscriptstyle(3)\!\!\!\:}\epsilon}
\def\@warning#1{\typeout{LaTeX Warning [l.\the\inputlineno]: #1.}}

\begin{document}


\input Titel
\input Einfuehrung
\input YamabeGleichung
\input NumericalImplementationYG
\input Zusammenfassung

\input biblio

\end{document}

%% file: Titel.tex
\title{Numerical Calculation of Conformally Smooth Hyperboloidal Data}

\author{Peter H\"ubner}
\email{pth@aei-potsdam.mpg.de}
\affiliation{%
  Max-Planck-Institut f\"ur Gravitationsphysik\\
  Albert-Einstein-Institut\\
  Am M\"uhlenberg 1\\
  D-14476 Golm\\
  FRG}

\begin{abstract}
This is the third paper in a series describing a numerical
implementation of the conformal Einstein equation.
This paper describes a scheme to calculate (three) dimensional data
for the  conformal field equations from a set of free functions.
The actual implementation depends on the topology of the spacetime.
We discuss the implementation and exemplary calculations for data
leading to spacetimes with one spherical null infinity (asymptotically
Minkowski) and for data leading to spacetimes with two toroidal
null infinities (asymptotically A3).
We also outline the (technical) modifications of the implementation
needed to calculate data for spacetimes with two and more spherical
null infinities (asymptotically Schwarzschild and asymptotically
multiple black holes).
\end{abstract}
%
%


\maketitle


%% file: Einfuehrung.tex
%
%
\section{Introduction}
\hspace{-\parindent}%
In the conformal approach to numerical relativity we give
data for the conformal field equations on a hyperboloidal initial
slice and calculate the conformal spacetime $(M,g_{ab})$ determined by
the data.
The region on which the conformal factor $\Omega$, which is one of our
variables, is positive can be identified with a physical spacetime
$(\tilde M,\tilde g_{ab})$ satisfying the Einstein equation.
The boundary $\{\Omega=0\}$ of this physical region represents
future null infinity (\Scri) and possibly future timelike infinity
($i^+$).
\\
Embedding the physical spacetime into a larger conformal spacetime
implies a number of advantages over conventional approaches:
Since null infinity is part of the grid, the determination of
gravitational radiation is a well-defined and a gauge-ambiguity-free
procedure.
Furthermore we avoid any influence of artificial boundaries onto the
result of our calculation, and we are enabled to use very efficient
high order discretisation techniques, reducing the required
computational resources by orders of magnitude.
\\
In the first two papers~\cite{Hu99ht,Hu99as} in the series we have
described the ideas behind the conformal approach, the general
mathematical background, important properties of the time evolution
equations, and the code used to integrate the time evolution
equations.
\\
To study properties of asymptotically flat solutions numerically, we
need to provide initial data, i.~e.\ we have to find solutions of the
conformal constraints.
In this paper we construct data for the conformal field equations by a
formally infinite order scheme.
The scheme consists of three basic elements, a Multigrid Newton 
Method to deal with the non-linearities, pseudo-spectral techniques to
achieve formally infinite order, and algebraic multigrid techniques to
invert the linearised elliptic operators.
\\
In section~\ref{YamabeEigen} we repeat the derivation of the Yamabe
equation by the Lichnerowicz ansatz and discuss those properties of the
solutions which are important for the numerical implementation by
analytical and numerical means.
In the next section we describe the strategy for calculating initial
data, details of the numerical implementation for the cases leading to
data for asymptotically Minkowski and asymptotically A3 spacetimes,
and give the results of two exemplary calculations.
In the last section we discuss how the scheme for the asymptotically
Minkowski data can be extended to data for multiple black hole
spacetimes by using well-established numerical techniques.
\\
To avoid too many repetitions, we assume the reader to be familiar
with the general approach as well as with the equations, both
discussed in part I and II of this series~\cite{Hu99ht,Hu99as}, and we
shall refer to equation~(n) of part~N by writing~(N/n).
To the interested reader we should also point out the
references~\cite{Hu93nu,Hu96mf,Fr98nta,Fr98ntb,Fr99ci} which describe 
a spherical symmetric (1D) and an axial symmetric (2D) implementation
of the conformal field equations.
There is also a recent review article by J.~Frauendiener in {\sl
  Living Reviews in Relativity}~\cite{Fr00ci}.
%
%

%% file: YamabeGleichung.tex
%
%
\section{The Yamabe equation and the smoothness of its solutions}
\label{YamabeEigen}
\hspace{-\parindent}%
A permissible set of data for the conformal time evolution equations
is given by a set of functions which satisfy the conformal
constraints.
Finding initial data is hence equivalent to finding a solution to the
conformal constraints.
\\
The constraints of the conformal field equation (I/14) 
are regular on the whole conformal spacetime $(M,g_{ab})$,
their principal part does not degenerate at \Scri{}.
Therefore, it would be nice, if we could solve these
constraints directly on our initial slice $\Sigma_{t_0}$.
At present we do not know how to do that.
The system of conformal constraints is a large coupled system of
equations and, with the exception of the case of spherical
symmetry~\cite{Hu96mf}, we do not know how to reduce it to a system to
which known numerical techniques can be applied, e.~g.\ a system of
elliptic equations.
\\
The way we construct initial data for the conformal field equations is
an indirect one, we first solve the constraints of the vacuum Einstein
equations and then construct the rest of the data from the conformal
constraints.
This amounts to providing a numerical implementation of the
approach which has been used in~\cite{AnCA92ot} by L.~Andersson,
P.~Chru\'sciel, and H.~Friedrich to prove existence of regular
data for the conformal field equations.
With this ansatz we are led to solving an elliptic equation of second
order.
A generalisation of the ansatz in~\cite{AnCA92ot} leads to a system of
four coupled equations which has been analysed in~\cite{AnC96so}.
We will restrict ourselves to the case of the ansatz
of~\cite{AnCA92ot} for simplicity.
In this case we already have the freedom to prescribe the conformal
metric on the initial slice.
Although this is not the full parameter space for the initial data, we
expect to have covered enough of the parameter space to study
interesting phenomenae.
\subsection{The Yamabe equation}
\hspace{-\parindent}%
Let $\tilde M$ be an asymptotically flat spacetime and $\tilde\Sigma$
a hyperboloidal submanifold, i.~e.\ a spacelike hypersurface extending
to future null infinity.
We denote by $\tilde h_{ab}$ the induced 3-metric, by ${\tilde
  k}_{ab}$ the induced second fundamental form, and by $\tilde k$ its
trace.
The Hamiltonian constraint then reads
\begin{equation}
\label{HamilConstr}
  \tRIII - {\tilde k}_{ab}{\tilde k}^{ab}
  + {\tilde k}^2  = 0,
\end{equation}
where $\tRIII$ is the Ricci scalar associated with $\tilde h_{ab}$.
If we denote by $\tDIII_a$ the covariant derivative induced by $\tilde
h_{ab}$, the vector constraint reads
\begin{equation}
\label{VekConstr}
  \tDIII^b {\tilde k}_{ab} - \tDIII_a \tilde k = 0.
\end{equation}
We want to construct initial data representing this geometric
situation.
To simplify our calculation we assume that on our slice 
\begin{subequations}
\label{LichnorowiczAnsatz}
\begin{equation}
  \label{tildek}
  {\tilde k}_{ab} = \frac{1}{3} {\tilde h}_{ab} \tilde k.
\end{equation}
The vector constraint~(\ref{VekConstr}) implies
\begin{equation}
  \label{tildeSpurk}
  \tilde k =  \const \ne 0.
\end{equation}
\end{subequations}
We assume 
\begin{equation}
  \tilde k > 0.
\end{equation}
Think of $\tilde\Sigma$ as being smoothly embedded into the conformal
extension of our spacetime through future null infinity $\Scri^+$ and
denote by $\bar\Sigma$ the closure of $\tilde\Sigma$ in this
extension and by ${\cal S}$ the boundary of $\tilde\Sigma$.
Let $\bar\Omega$ be a boundary defining function with
non-vanishing gradient on ${\cal S}$, positive in the
interior \footnote{The choice of the sign is arbritrary, since
  the conformal and the physical metric are connected by the square of
  the conformal factor.} and vanishing on ${\cal S}$, and let $h_{ab}$ be
an (almost) arbritrary metric on $\bar\Sigma$. 
\\
To reduce the Hamiltonian constraint~(\ref{HamilConstr}) to an
elliptic equation for a scalar function $\phi$, we make the so-called
Lichnerowicz ansatz, which reads in our case
\begin{eqnarray}
  \label{tildeh}
  {\tilde h}_{ab} 
    & = & \left( \frac{\bar\Omega}{\phi^2} \right)^{-2} h_{ab}.
\end{eqnarray}
With this ansatz ${\tilde h}_{ab}$ is singular at ${\cal S}$
indicating that ${\cal S}$ represents an infinity.
\\
The Hamiltonian constraint becomes the so-called Yamabe equation,
\begin{eqnarray}
\label{Yamabe}
  & &
  4 \, \bar\Omega^2 \DeIII \phi
  - 4 \, \bar\Omega (\!\DIII^a \bar\Omega)(\!\DIII_a \phi) \nonumber\\
  & & \qquad
  - \left( \frac{1}{2} \RIII \, \bar\Omega^2 + 2 \bar\Omega \DeIII\bar\Omega
           - 3 (\!\DIII^a \bar\Omega) (\!\DIII_a \bar\Omega) 
    \right) \phi
  - \frac{1}{3} {\tilde k}^2 \phi^5
  = 0,
\end{eqnarray}
where $\DIII_a$, $\DeIII$, and $\RIII$ are the covariant derivative,
the Laplace operator, and the Ricci scalar associated with $h_{ab}$.
The Yamabe equation is the equation, which we are going to solve
numerically.
It is an ``elliptic'' equation with a principal part which vanishes at
the boundary.
\\
After having solved the Yamabe equation~(\ref{Yamabe}) the remaining
members of a minimal set of data which consists of
$(h_{ab},\Omega,k_{ab},\Omega_0)$ (cf.~II) are given by
\begin{eqnarray}
\label{Om}
  \Omega & = & \frac{\bar\Omega}{\phi^2} \\
\label{extrCurv}
  k_{ab} & = & \frac{1}{3} k h_{ab} \\
\label{OmO}
  \Omega_0 & = &
  \frac{1}{3} \left( \Omega k - \tilde k \right),
\end{eqnarray}
where $k$ is an arbritrary function.
Its choice is pure conformal gauge, it does not influence the
physical spacetime obtained~(cf.\ also~\cite[section 3]{Fr83cp}
or~\cite[subsection 2.1.1]{Hu95gr}).
\\
To calculate a complete set of data for the conformal field equations
from a minimal set of data we use the conformal constraints to
calculate $\gamma^a{}_{bc}$, $\ROI_a$, $\RII_a{}^a$, $\Omega_a$,
$\omega$, $f_{\RII\,ab} := \Omega \RII_{ab}$, $f_{B\,ab}:= \Omega
B_{ab}$, and $f_{E\,ab} = \Omega E_{ab} - 1/2 \RII_{ab}$ (cf.~II for
more details).
\\
Due to our assumption (\ref{tildek}) the magnetic part of the Weyl
tensor $B_{ab}$ vanishes.
For the case of vanishing conformal extrinsic curvature, i.~e.\ $k=0$,
the quantity $\ROI_a$ also vanishes.
\\
The following theorem by L.~Andersson, P.~Chru\'sciel and
H.~Friedrich~\cite{AnCA92ot} gives existence and uniqueness of
positive solutions to the Yamabe equation and guarantees regularity of
$\RII_{ab}$ and $E_{ab}$:
\newtheorem{theorem}{Theorem}
\begin{theorem}[Andersson, Chru\'sciel, Friedrich]
\label{SatzACF}
Suppose $(\bar\Sigma_{t_0},h_{ab},\bar\Omega)$ is a 3-dimensional, orientable,
compact, smooth Riemann space with boundary~${\cal S}$, $\bar\Omega=0$
on ${\cal S}$ and positive elsewhere. 
Then there exists a unique positive solution $\phi$ of
equation~(\ref{Yamabe}) and the following conditions are equivalent:
\begin{enumerate}
\item The function $\phi$ as well as the corresponding complete set of
  data, determined on the interior $\tilde\Sigma_{t_0}$ of
  $\bar\Sigma_{t_0}$ , extend smoothly to all of
  $(\bar\Sigma_{t_0},h_{ab})$.
\item The electric part of the conformal Weyl tensor $\Omega
  E_{ab}$ goes to zero at~${\cal S}$.
\item The extrinsic 2-curvature induced by $h_{ab}$ on~${\cal S}$ is
  pure trace.
\end{enumerate}
\end{theorem}
If we did not require, that the extrinsic 2-curvature of~${\cal S}$
were pure trace, $f_{\RII\,ab}$ and $f_{E\,ab}$ would not be vanishing
on the boundary and therefore $\RII_{ab}$ and $E_{ab}$ would not be
regular there.
For this reason we choose the conformal 3-metric in such a way that
the tracefree part of the extrinsic 2-curvature of~${\cal S}$ with
respect to $h_{ab}$ vanishes.
\\
Observe that if we want to have a regular positive solution $\phi$ of
(\ref{Yamabe}) we cannot specify boundary values for $\phi$ on~${\cal
  S}$.
Since for a regular solution the principle part vanishes at the
boundary, we must have
\begin{equation}
  \label{YamabeBoundaryValues}
  \left.(\nabla^a \bar{\Omega}) (\nabla_a \bar{\Omega})
  - \frac{{\tilde k}^2}{9} \phi^4 \right|_{\cal S}= 0.  
\end{equation}
\subsection{Smoothness of solutions of the Yamabe equation on the
  initial slice}
\begin{figure}[htbp]
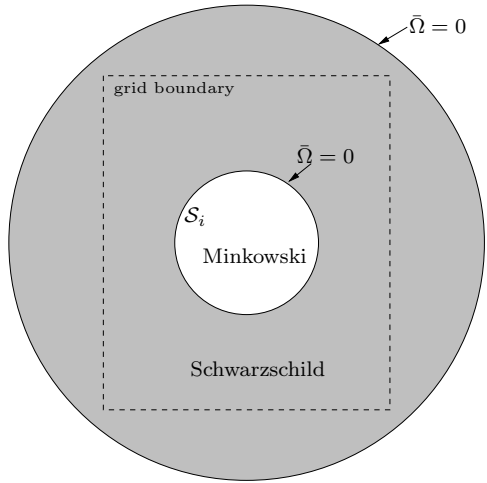

  \begin{center}
    \begin{minipage}[t]{7cm}
      \input SchwMinkArg.pstex_t
      \vskip0.5em
      \caption{\label{SchwMinkArg}A Kruskal-Schwarzschild slice
        enclosing a Minkowski slice.}
    \end{minipage}
  \end{center}
\end{figure}

\begin{figure}[htbp]
  \begin{center}
    \begin{minipage}[t]{10cm}
      \includegraphics[width=10cm]{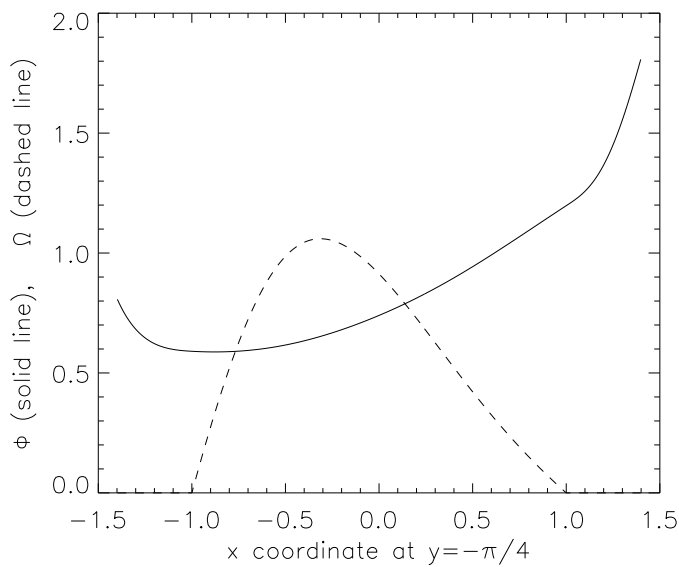}%
      \vskip0.5em
      \caption{\label{SprungPhi}Plot of $\phi$ (solid line) and
        $\Omega$ (dashed line) along the $y=\pi/4$ line.}
    \end{minipage}
  \end{center}
\end{figure}

\begin{figure}[htbp]
  \begin{center}
    \begin{minipage}[t]{10cm}
      \includegraphics[width=10cm]{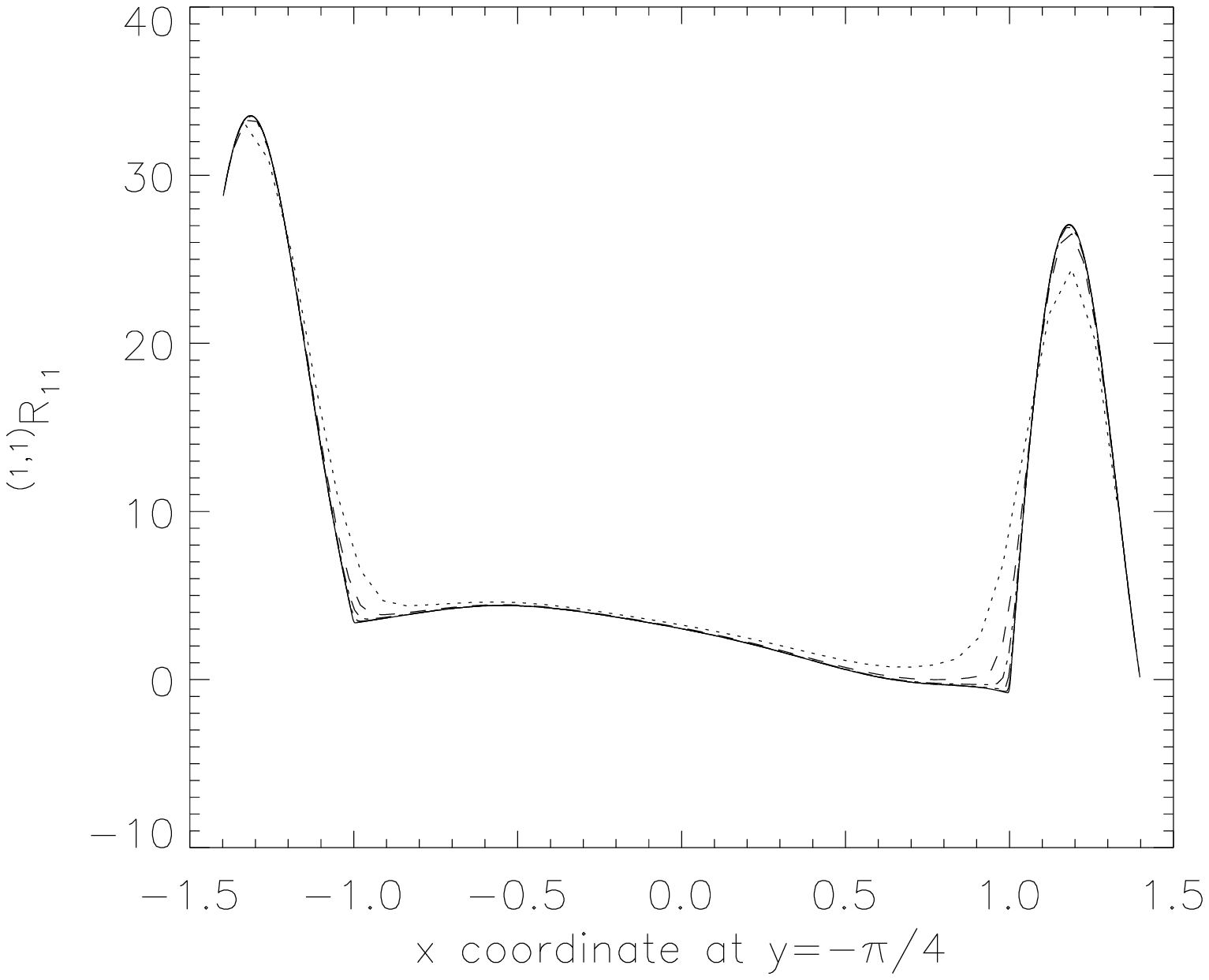}%
      \vskip0.5em
      \caption{\label{SprungR11}$\RII_{\underline{x}\underline{x}}$
        for grids from $20 \times 20$ (dotted line) to 
        $640 \times 640$ (solid line).}
    \end{minipage}
  \end{center}
\end{figure}

\begin{figure}[htbp]
  \begin{center}
    \begin{minipage}[t]{10cm}
      \includegraphics[width=10cm]{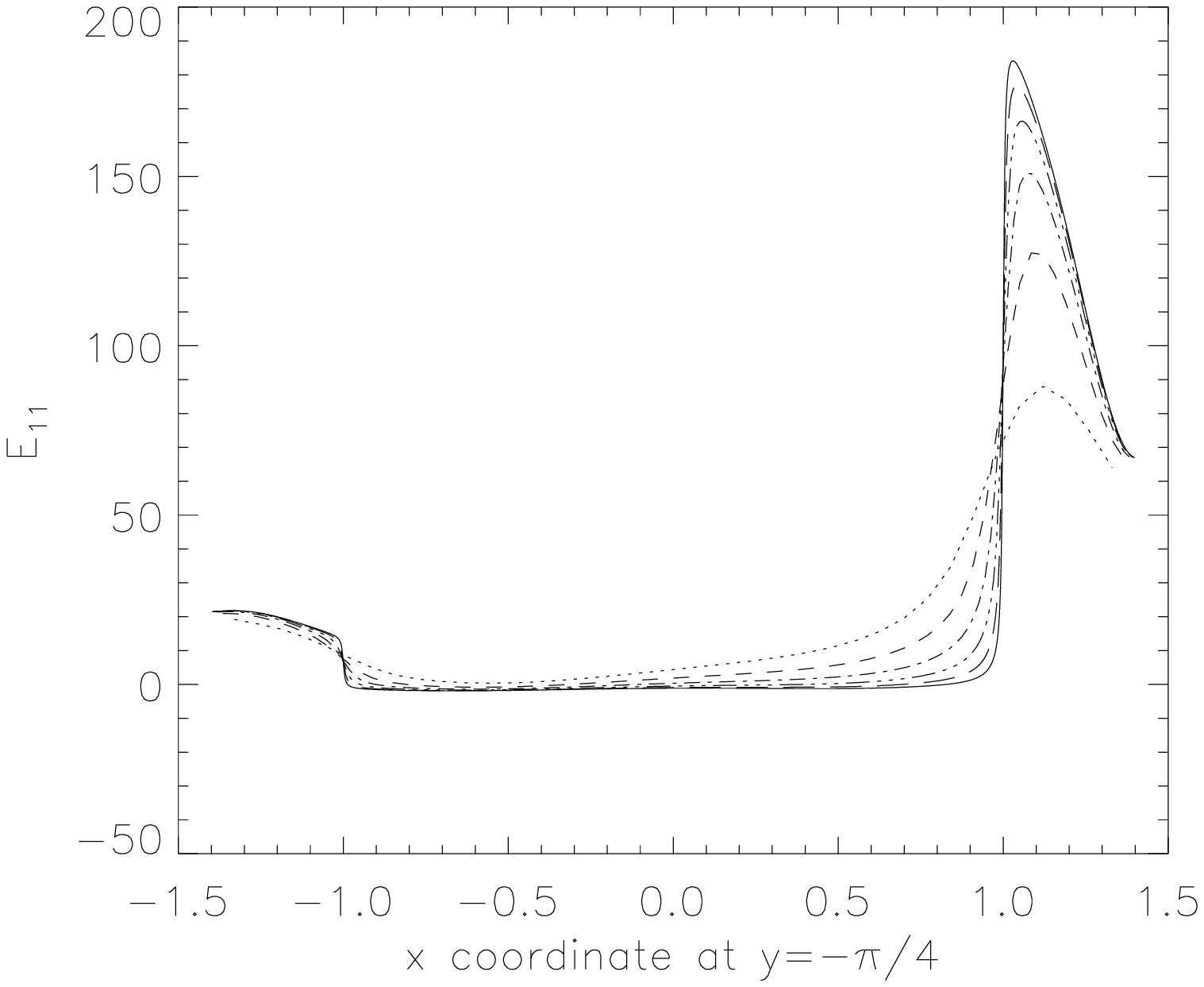}
      \vskip0.5em
      \caption{\label{SprungE11}$E_{\underline{x}\underline{x}}$ for
        grids from $20 \times 20$ (dotted line) to $640 \times 640$
        (solid line).}
    \end{minipage}
  \end{center}
\end{figure}

\hspace{-\parindent}%
In our setup $\bar\Sigma_{t_0}$ will be a true subset of
the initial slice $\Sigma_{t_0}$ represented by the grid.
To construct a complete set of data on $\Sigma_{t_0}$ one is
tempted to give $\bar\Omega$ and $h_{ab}$ on $\Sigma_{t_0}$ with
$\bar\Omega \le 0$ outside $\bar\Sigma_{t_0}$ and solve (\ref{Yamabe})
on $\Sigma_{t_0}$.
In the general case we obtain a complete set of data which is not
continuous on~${\cal S}$.
Since our discretisation of the time integrator requires sufficiently
smooth data we cannot proceed this way.
In the following we will give an argument, why we have to expect a
non-continuity, and we will show a numerical example.
\\[\baselineskip]
To calculate the curvature variables $\RII_{ab}$ and $E_{ab}$ from
$f_{\RII\,ab}$ and $f_{E\,ab}$ we have to divide by $\Omega$.
To calculate the limit for $\Omega\rightarrow 0$, which is the value
at~${\cal S}$, we apply l'Hopital's rule and get
\begin{equation}
  \label{LHopital}
  \left.\frac{f}{\Omega}\right|_{\Omega=0} = 
  \frac{\DIII^a\Omega \DIII_a f}{\DIII^a\Omega \DIII_a \Omega}.
\end{equation}
For the highest derivatives of $\phi$ which appear in the expressions
of $\RII_{ab}$ and $E_{ab}$ we find
\begin{subequations}
  \begin{eqnarray}
    \RII_{ab} & \sim & \partial^3\phi, \\
    E_{ab} & \sim & \partial^4\phi \,.
  \end{eqnarray}
\end{subequations}
\\
Let us now look at the following simple situation of figure~\ref{SchwMinkArg}.
Suppose we have given a spherical symmetric $\bar\Omega$, a spherical
symmetric $h_{ab}$ such that~${\cal S}_i$ is the unit 
sphere, and $k=0$.
Suppose $\bar\Omega$ is positive in the part labeled ``Minkowski'',
which has one boundary diffeomorphic to $S^2$, and negative in the
part labeled ``Schwarzschild'', which has two boundary components of
topology $S^2$. 
Then, due to Birkhoff's theorem, $\{\bar\Omega>0\}$ is a
hyperboloidal slice of Minkowski spacetime, and $\{\bar\Omega<0\}$ is
a hyperboloidal slice of a Schwarzschild spacetime, since there are no
spacelike slices in Minkowski space connecting two null infinities.
The latter must have mass $m\ne 0$, whereas the mass of the Minkowski
part necessarily vanishes.
The Bondi mass of the initial slice is given by the integral
over~${\cal S}$ with an integrand which is a polynomial expression of
our variables involving components of $E_{ab}$~\cite{HuWXXxx}.
Since the mass is $0$, if we take the integral on the interior side,
and non-vanishing, if we take the integral on the exterior side, the
integrand cannot be continuous.
The only source of the non-continuity can be the derivatives of
$\phi$. 
From the proof of theorem~\ref{SatzACF} it follows that $\phi$ is
at least $C^3$ across the boundary~${\cal S}$.
Since the highest derivatives of $\phi$ in $E_{ab}$ are of fourth
order, $\phi$ can in general not be $C^4$ across the boundary~${\cal
  S}$.
\\
In a numerical code we would, of course, give boundary values for
$\phi$ on the grid boundaries instead of specifying where the
outer~${\cal S}$ is placed.
However, unless we happen to prescribe by coincidence the appropriate
boundary data on the grid boundary, we can expect to get the same
behaviour.
To check this hypothesis we have performed numerical test calculations
for asymptotically A3 data with one Killing vector (a 2D calculation).
\\
In the following we give the results of a typical numerical
calculation. 
In this 2D calculation we have given the same free functions as
in~\cite{Fr99ci},
\begin{subequations}
  \begin{eqnarray}
    \bar\Omega & = & \frac{1}{2} \left( 1 - x^2 \right)              \\
    h_{xx}     & = & 2 e^{- 2 x \cos(y) }                            \\
    h_{xy}     & = & 2 \bar\Omega \left( x^2 -\sin(y^2) \right) 
                     e^{- x \cos(y) }                                \\
    h_{yy}     & = & 2 \left( 1 + \bar\Omega^2 
                     \left( x^2 -\sin(y^2) \right)^2 \right)           \\
    h_{zz}     & = & 2 e^{- 2 x \cos(y) }               \\
    k          & = & 0,
  \end{eqnarray}
\end{subequations}
where $(x,y) \in [-1.4,1.4] \times [-\pi,\pi]$.
The free functions satisfy the regularity condition on~${\cal S}$.
We then solve the Yamabe equation by discretising it by centered 
second order stencils, inverting the resulting sparse matrix with
the AMG library~\cite{St99am,RuS87am}, and taking care of the
non-linearities by a Multigrid Newton Method~\cite{Br97FE}.
\\
Although the free functions depend on both coordinates, we
plot the results only for $y=-\pi/4$ for the reason of simpler graphics.
The solution $\phi$ (solid line) for the calculation with the 
highest resolution performed, which is a $640 \times  640$ grid
calculation, is shown in figure~\ref{SprungPhi}.
We have also plotted the conformal factor $\Omega$ (dashed line)
to indicate the location of the~${\cal S}$s at $x=-1$ and $x=1$.
The function $\phi$ seems to be perfectly smooth across
the~${\cal S}$s.
\\
By calculating the quantities $\RII_{ab}$ and $E_{ab}$ we see that
this is not the case.
To do so we use the second order scheme described in II.
Figure~\ref{SprungR11} shows $\RII_{\underline{x}\underline{x}}$ for
the grid sizes of $20 \times 20$, $40 \times 40$, $80 \times 80$, 
$160 \times 160$, $320 \times 320$, and $640 \times 640$ gridpoints
(by $\RII_{\underline{x}\underline{x}}$ we denote the $x$ coordinate
components of the tensor $\RII_{ab}$).
Increasing line density correlates with finer grids.
Obviously convergence near the two~${\cal S}$s is slow --- the difference
between the various grids is larger --- and obviously 
$\RII_{\underline{x}\underline{x}}$ converges against a function which
is only $C^0$ at the two~${\cal S}$s.
This non-smoothness also explains the slow convergence rate.
\\
Convergence is even slower in figure~\ref{SprungE11}, where we plot
$E_{\underline{x}\underline{x}}$.
The solution slowly converges to a function with steps at
the two~${\cal S}$s as expected from the theoretical considerations in
the first part of the subsection.
%
%

%% file: SchwMinkArg.pstex_t
\begin{picture}(0,0)%
\epsfig{file=SchwMinkArg.pstex}%
\end{picture}%
\setlength{\unitlength}{0.00041700in}%
\begingroup\makeatletter\ifx\SetFigFont\undefined%
\gdef\SetFigFont#1#2#3#4#5{%
  \reset@font\fontsize{#1}{#2pt}%
  \fontfamily{#3}\fontseries{#4}\fontshape{#5}%
  \selectfont}%
\fi\endgroup%
\begin{picture}(5990,5990)(2106,-6656)
\put(4576,-3886){\makebox(0,0)[lb]{\smash{\SetFigFont{8}{9.6}{\familydefault}{\mddefault}{\updefault}Minkowski}}}
\put(4351,-3361){\makebox(0,0)[lb]{\smash{\SetFigFont{8}{9.6}{\familydefault}{\mddefault}{\updefault}${\cal S}_i$}}}
\put(3451,-1711){\makebox(0,0)[lb]{\smash{\SetFigFont{6}{7.2}{\familydefault}{\mddefault}{\updefault}grid boundary}}}
\put(4426,-5311){\makebox(0,0)[lb]{\smash{\SetFigFont{8}{9.6}{\familydefault}{\mddefault}{\updefault}Schwarzschild}}}
\put(7201,-961){\makebox(0,0)[lb]{\smash{\SetFigFont{8}{9.6}{\familydefault}{\mddefault}{\updefault}$\bar\Omega=0$}}}
\put(5776,-2611){\makebox(0,0)[lb]{\smash{\SetFigFont{8}{9.6}{\familydefault}{\mddefault}{\updefault}$\bar\Omega=0$}}}
\end{picture}

%% file: NumericalImplementationYG.tex
%
%
\section{Calculating extended hyperboloidal initial data --- the
  numerical implementation}
\subsection{A rough description of the main ideas}
\hspace{-\parindent}%
What has been said in the previous section suggests the following
strategy to calculate extended hyperboloidal initial data:
\begin{enumerate}
\item\label{SolveYamabe}Solve the Yamabe equation~(\ref{Yamabe}) on
  $\bar\Sigma_{t_0}$ to get a minimal set of data.
\item\label{Min2Full}Calculate a complete set of data on $\bar\Sigma_{t_0}$.
\item\label{Extend}Smoothly extend the data from $\bar\Sigma_{t_0}$ to
  $\Sigma_{t_0}$.
\end{enumerate}
The main part of the numerical implementation of
step~\ref{SolveYamabe} and step~\ref{Min2Full} is solving ``elliptic''
equations with a principal part which degenerates at the boundary.
To be able to obtain sufficient accuracy with an acceptable
consumption of computer resources even without symmetry assumptions
(3D) we use pseudo-spectral methods similar to the 2D calculations
described in~\cite{Fr99ci}.
Since there exists a huge amount of literature on pseudo-spectral
methods, we only very briefly describe our choices.
The reader who wants to learn about the basics is referred to the
literature, e.~g.~\cite{QuV97NA} for the general theory
or~\cite{BoFE96sm} for other general relativistic applications.
\\
Since pseudo-spectral methods are formally infinite order, the grid on
which we discretise and invert the elliptic operator can be pretty
coarse.
The accuracy achieved is very high, as soon as the spectral grid
resolution is sufficient to represent the structure of the solution.
The limit of achievable accuracy is given by the accumulation of rounding
errors in the fast Fourier transformation, the matrix inversion, and
the consecutive solving of elliptic equations.
Typically, using $32$ gridpoints in each dimension is sufficient to
achieve an accuracy which cannot be further improved.
\\
For step~\ref{Extend} we use the representation of our variables
in the spectral basis to calculate a smooth extension.
\subsection{The Yamabe solver and the $f/\Omega$ divider}
\hspace{-\parindent}\label{YamabeSolver}%
For simplicity we write the Yamabe equation in the form
\begin{equation}
  \label{FormalYamabe}
  {\bf E} \, \phi - \frac{1}{3} {\tilde k}^2 \phi^5 = 0,
\end{equation}
where ${\bf E}$ is a linear operator.
\\
To deal with the nonlinearity we use the Multigrid Newton Method for
nonlinear systems described in~\cite{Br97FE}:
\\
Assuming we have given an approximate solution $\phi^{(n)}$ we
calculate a hopefully better solution
\begin{equation}
  \phi^{(n+1)} = \phi^{(n)} + \epsilon \> \delta\phi,
\end{equation}
where the correction is the solution of 
\begin{equation}
  {\bf E}_l \delta\phi :=
  {\bf E} \delta\phi - 5 \left(\phi^{(n)}\right)^4 \delta\phi
  =
  - {\bf E} \phi^{(n)} + \left(\phi^{(n)}\right)^5,
\end{equation}
which is obtained by linearising the nonlinear
equation~(\ref{FormalYamabe}).
Since $\delta\phi$ is a correction to the solution, the order of the
scheme and the eventual quality of the approximation depend on the order
and the accuracy of the calculation of the residuum 
$ - {\bf E} \phi^{(n)} + \left(\phi^{(n)}\right)^5$, but not on the
discretisation of ${\bf E}_l$.
\\
The value of the parameter $\epsilon$ is determined in the Multigrid
Newton Method~\cite{Br97FE} by the following algorithm:
We take the first value of $\epsilon$ in the sequence
$1,\frac{1}{2},\frac{1}{4},\ldots$ for which
\begin{equation}
 \| {\bf E} \phi^{(n+1)} - \left(\phi^{(n+1)}\right)^5 \|
 \le
 \left( 1 - \frac{\epsilon}{2} \right) 
 \| {\bf E} \phi^{(n)} - \left(\phi^{(n)}\right)^5 \|
\end{equation}
for an appropriate norm $\|.\|$, in our case the $L_2$ norm.
This adaption of $\epsilon$ on each step stabilises the scheme and
makes it less sensitive to a good initial guess $\phi^{(0)}$.
Numerical experiments with a fixed $\epsilon$ were only stable for
$\epsilon<1/2^n$, where $n$ is the dimension of the hypersurface
$\bar\Sigma_{t_0}$.
\\
The residuum is calculated by pseudo-spectral methods, which means, that
from the values of $\phi^{(n)}$ at the gridpoints we calculate the
spectral coefficients.
Then we differentiate in the spectral space, where differentiation is
essentially a multiplication, and transform back to the grid to get
high order approximations of the derivatives.
With these values of the derivatives we calculate the value of the
residuum on the gridpoints.
\\
The order of the differentiation with respect to a coordinate $x^a$ is
approximately the number of functions in the spectral basis for
this coordinate, which is approximately the number of gridpoints in the
corresponding dimension (a more accurate statement depends on the spectral
basis, see below). 
Therefore the order of the differentiation increases with the number
of gridpoints.
This increase of the order with the number of gridpoints is called
``formally infinite order''.
\\
For reasons of numerical efficiency the choice of the spectral basis
is restricted to those bases for which we can use fast Fourier
transformation (FFT) techniques to perform the transformation to and
from the spectral space.
\\
To perform the FFTs we use the FFTW library by M.~Frigo and
S.~G.~Johnson, which is a C subroutine library for computing the
discrete Fourier transform (DFT) in one or more dimensions, of both
real and complex data, and of arbitrary input size~\cite{FrJ98fa}.
The library includes a parallel version for POSIX threads and the FFT
is done by an $N \log N$ algorithm even if the prime factor
decomposition of $N$ contains large prime numbers. 
This has large advantages over simple $N \log N$ algorithms which are
based on the assumption that the number of gridpoints N is a power of
$2$, e.~g.\ when it comes to the issue of grid refinements in~3D.
\\
The derivatives in the operator ${\bf E_l}$ are discretised by three
point stencils.
These stencils depend on the spacing of the grid which is given by the
choice of the spectral basis.
We are going to describe it below when we discuss the specific cases.
Due to the dependence on the spectral basis we call this grid a spectral
grid.
In general the spectral grid does not coincide with the grid used for
time evolution and is significantly coarser.
\\
Where the boundary of the spectral grid coincides with ${\cal S}$, the
initial approximation $\phi^{(0)}$ for the solution is chosen such
that it coincides with (\ref{YamabeBoundaryValues}) on ${\cal S}$.
Then the boundary value for $\delta\phi$ is $0$ in each iteration.
If the grid boundary is inside physical spacetime, which is the case
for more complicated forms of $\bar\Sigma_{t_0}$, where we need
overlapping spectral grids to cover the physical part of the initial
slice (multiple black holes, subsection~\ref{dataAsympnBH}), the
boundary values for $\delta\phi$ are deduced from the other grids and
iterated in a Schwarz alternating procedure.
\\
After having provided boundary values we use algebraic multigrid
techniques to solve for $\delta\phi$.
To do so we use the AMG library~\cite{St99am,RuS87am}, which was
kindly provided to us by K.~St\"uben from the Gesellschaft f\"ur
Mathematik und Datenverarbeitung.
Different topologies of the spaces on which we solve the elliptic
equations show up in different structures of the discretisation
matrix of ${\bf E_l}$.
Since AMG automatically derives a multigrid coarsening strategy from
the structure of the discretisation matrix, only minor changes need to
be made in the code which inverts the elliptic operator to adapt it to
the various cases described below.
This is a huge advantage.
The price to pay is the fact, that the AMG library is the only part of
the code which does not give excellent parallel scaling, in fact the
AMG version used does not run in parallel at all.
For a typical 3D run, which produces data for a time evolution
grid of $100^3$ or more gridpoints, the total time spent in the AMG
library is of order of a few percent of the time spent in the rest of
the code, although the rest has excellent parallel performance.
Therefore, the non-scaling of the AMG part of the code is not a
serious issue.
\\[\baselineskip]
To calculate $g=f/\Omega$ we use the same numerical techniques applied to
equation~(II/7), which can formally be written as
\begin{equation}
  {\bf G} g = F[f]
\end{equation}
and has the ``linearised'' form
\begin{equation}
  {\bf G}_l \delta g =
    - {\bf G} g^{(n)} + F[f].
\end{equation}
Observe that the metric used in (II/7) does not need to be identical
with the 3-metric $h_{ab}$.
And indeed, the code becomes much simpler if we use the diagonal
Euclidean 3-metric $\delta_{ab}$.
\\
As boundary values we use (II/8) on the grid boundaries which coincide
with ${\cal S}$ and $f/\Omega$ on the grid boundaries which lie in the
physical part.
Of course, due to the latter, the grid boundaries which lie in the
physical part must not intersect ${\cal S}$.
As initial approximation we use a low order approximation of
equation~(II/7).
\\
The attentive reader may ask the following.
Equation~(II/7) is a linear equation for $g$, what is the reason for
doing an iteration which is typically used to deal with nonlinear
terms.
\\
The reason is very simple.
If we discretised ${\bf G}$ to low order, we would get a sparse
discretisation matrix, but also an inaccurate solution.
If we discretised with high order stencils, we would get a
not-very--sparse discretisation matrix, which is difficult to invert,
and a complicated boundary treatment.
By using the iteration on the linear equation we get a formally
infinite order solution, which turns out to be very accurate, but
we still have a sparse discretisation matrix.
The repeated inversion of ${\bf G}_l$ is the price to pay.
\subsection{The case of asymptotically A3 spacetimes}
\hspace{-\parindent}\label{dataAsympA3}%
From the viewpoint of the initial data solver the simplest class of
initial data are asymptotically A3 spacetimes.
There the physical part $\tilde\Sigma_{t_0}$ of an initial slice has
topology $R\times T^2$, and the boundary ${\cal S}$ consists of two 
two-dimensional tori $T^2$~(figure 4 of I).
By doing an appropriate coordinate transformation within the initial
data surface we can always ensure, that the coordinates $y$ and
$z$ are $2\pi$ periodic and parametrise the two-dimensional torus
\footnote{We describe the 3D code here. We also have a 2D code,
  where we made use of the simplifications implied by an hypersurface
  orthogonal Killing vector $\partial_z{}^a$. From the description of
  the 3D code, the 2D code should be obvious.}, and that the
two parts of the boundary ${\cal S}$, in the following called  ${\cal
  S}_l$ and ${\cal S}_r$, are at constant $x$ values $x=a$ and $x=b$.
Although the code allows arbritrary values of $a$ and $b$, as well as
other periodicities in the $y$ and $z$ direction, we only describe the
case with $a=-1$, $b=1$, and $2\pi$ periodicity for simplicity.
\\
Since $y$ and $z$ are $2\pi$ periodic we can simply use the Fourier
functions $e^{i\,k_y\,y}$ and $e^{i\,k_z\,z}$ as basis in the $y$ and
$z$ direction.
The spacing of the spectral grid in $y$ and $z$ direction is
equidistant, the $N_{y,z}$ nodes of the spectral grid are placed at
\begin{subequations}
  \label{FourierPoints}
  \begin{eqnarray}
    y_j & := & \frac{2\pi j}{N_y}  \\
    z_k & := & \frac{2\pi k}{N_z}
  \end{eqnarray}
\end{subequations}
\\
In the $x$ direction we use Chebychev polynomials 
\begin{equation}
  \label{TschPoly}
  T_{k_x}(x) := \cos\left( k_x \arccos(x) \right).
\end{equation}
The gridpoints are placed at the Gauss-Lobatto nodes
\begin{equation}
  \label{GaussLobatto}
  x_i := \cos \frac{\pi i}{N_x},
\end{equation}
where $N_x + 1$ is the number of gridpoints in the $x$ direction.
\\
For every function
\begin{equation}
  \label{OrigTscheb}
  f: [-1,1] \rightarrow R,
\end{equation}
there is a one-to-one relation to the function 
\begin{equation}
  \label{CosTscheb}
  f \circ cos : [0,\pi] \rightarrow R.
\end{equation}
We can reduce the Chebychev transformation to a Fourier
transformation, which of course can be treated by FFT techniques, by
transferring the non-equidistant gridpoints in $[-1,1]$ to
equidistant gridpoints in $[0,\pi]$ and extending the range to the
interval $[0,2\pi[$ by applying the symmetry 
\begin{equation}
  \label{CosSym}
  \cos(\pi+x)=-\cos(x).
\end{equation}
For a function $f(x,y,z)$ the spectral representation 
\begin{equation}
  \label{SpektralSumme}
  f(x,y,z) := 
  \sum_{k_x=0}^{N_x} \quad \sum_{k_y=-N_y/2}^{N_y/2} 
  \quad \sum_{k_z=-N_z/2}^{N_z/2}
    a_{k_x k_y k_z} T_{k_x}(x) e^{i\,k_y\,y} e^{i\,k_z\,z}
\end{equation}
is given by the coefficients $a_{k_x k_y k_z}$. Depending on $N_y$ and
$N_z$ some of the coefficients are zero and due to the reality of $f$ the
complex numbers $a_{k_x k_y k_z}$ are not independent.
\\
Then the coefficients $a^{xyz}{}_{k_x k_y k_z}$ of the derivatives
$\partial_{xyz}$ are given by
\begin{subequations}
  \begin{eqnarray}
    a^y{}_{k_x k_y k_z} & = & k_y \, a_{k_x k_y k_z} \\
    a^z{}_{k_x k_y k_z} & = & k_z \, a_{k_x k_y k_z}
  \end{eqnarray}
and the recursion relation 
\begin{equation}
  \label{KoeffDerivTscheb}
  \left\{ 
    \begin{array}{l}
      a^x{}_{N_x k_y k_z} = a^x{}_{N_x+1 \: k_y k_z} = 0 \\
      c_{k_x} a^x{}_{k_x k_y k_z} = a^x{}_{k_x+2 \: k_y k_z} + 2 (k_x+1)
        a^x{}_{k_x+1 \: k_y k_z}, \quad k_x = N_x-1, \ldots, 0, 
    \end{array}
  \right.
\end{equation}
\end{subequations}
where $c_0=2$ and $c_{k_x}=1$ for $k_x \ge 1$.
\\
Due to (\ref{KoeffDerivTscheb}) we loose one discretisation order with
each $\partial_x$.
\\
There exist FFT algorithms which take into account the even symmetries
of the function~(\ref{CosTscheb}), the so-called fast cosine
transformations~\cite{Lo92CF}.
They are significantly more complicated than a normal FFT, but reduce
the required space by a factor of $2$ approximately.
Since the space temporarily required in the initial data solver part
of the code is significantly less than the space temporarily required
in the time evolution part, we abstained from making use of the fast
cosine transformation techniques and the implied memory savings.
\\
Therefore, to calculate the spectral representation on the spectral
grid, we proceed as follows: We first relabel the grid covering 
$[-1,1] \times [0,2\pi[ \times [0,2\pi[$ to equidistantly cover 
$[0,\pi] \times [0,2\pi[ \times [0,2\pi[$ and then extend it to 
$[0,2\pi[ \times [0,2\pi[ \times [0,2\pi[$.
We use a three dimensional FFT for real numbers on the extended grid
to calculate the spectral representation.
Although our scheme can deal with any number of spectral gridpoints
$(N_x,N_y,N_z) \ge (2,2,2)$, for reasons of numerical efficiency it is
advisable to use values for which we get efficient FFTs.
Of course, the large grid covering 
$[0,2\pi[ \times [0,2\pi[ \times [0,2\pi[$ is only used for
calculating the spectral transformation and its inverse, tasks like
calculating the residuum and the inversion of the discretisation
matrix of ${\bf E}_l$ are performed  on the original, smaller grid
covering $[-1,1] \times [0,2\pi[ \times [0,2\pi[$.
The procedure for the inverse transformation is obvious from what has
been said.
\\
To derive the discretisation matrix of ${\bf E}_l$ we discretise the
derivatives in ${\bf E}_l$ as follows
\begin{subequations}
  \label{ElDiskretisierung}
  \begin{eqnarray}
  \partial_x f & \rightarrow & 
    - \frac{x_{i+1}-x_i}%
           {\left(x_{i+1}-x_{i-1}\right)\left(x_{i}-x_{i-1}\right)}
      f_{i-1,j,k}
    + \frac{\left(x_{i+1}-x_i\right)-\left(x_i-x_{i-1}\right)}%
           {\left(x_{i+1}-x_i\right)\left(x_{i}-x_{i-1}\right)}
      f_{i,j,k} \nonumber \\
    & & {} 
    + \frac{x_i-x_{i-1}}%
           {\left(x_{i+1}-x_{i-1}\right)\left(x_{i+1}-x_i\right)}
      f_{i+1,j,k}
  \\
  \partial_y f & \rightarrow &
    - \frac{1}{y_{j+1}-y_{j-1}} f_{i,j-1,k}
    + \frac{1}{y_{j+1}-y_{j-1}} f_{i,j+1,k} \\
  \partial_z f & \rightarrow &
    - \frac{1}{z_{k+1}-z_{k-1}} f_{i,j,k-1}
    + \frac{1}{z_{k+1}-z_{k-1}} f_{i,j,k+1} \\
  \partial^2_x f & \rightarrow & 
    \frac{2}{\left(x_{i+1}-x_{i-1}\right)\left(x_{i}-x_{i-1}\right)}
      f_{i-1,j,k}
    - \frac{2}{\left(x_{i+1}-x_i\right)\left(x_{i}-x_{i-1}\right)}
      f_{i,j,k} \nonumber \\
    & & {}
    + \frac{2}{\left(x_{i+1}-x_{i-1}\right)\left(x_{i+1}-x_i\right)}
      f_{i+1,j,k}
  \\
  \partial^2_y f & \rightarrow & 
    \frac{4}{\left(y_{j+1}-y_{j-1}\right)^2} f_{i,j-1,k}
    - \frac{8}{\left(y_{j+1}-y_{j-1}\right)^2} f_{i,j,k}
    + \frac{4}{\left(y_{j+1}-y_{j-1}\right)^2} f_{i,j+1,k}
  \\
  \partial^2_z f & \rightarrow &  
    \frac{4}{\left(z_{k+1}-z_{k-1}\right)^2} f_{i,j,k-1}
    - \frac{8}{\left(z_{k+1}-z_{k-1}\right)^2} f_{i,j,k}
    + \frac{4}{\left(z_{k+1}-z_{k-1}\right)^2} f_{i,j,k+1} \,.
  \end{eqnarray}
\end{subequations}
The mixed second derivatives are the obvious combinations of the first
derivatives.
\\
Combining the previous with subsection~\ref{YamabeSolver} we have a
very efficient elliptic solver for the Yamabe equation
(cf.~subsection~\ref{TwoExampl} for numbers).
The solution of the Yamabe equation determines a minimal set of data.
\\
Next we calculate the first and second derivatives of the minimal set
of data by pseudo-spectral methods, which immediately leads to values
for $\gamma^a{}_{bc}$, $\ROI_a$, $\RII_a{}^a$, $\Omega_a$, 
$\omega$, $f_{\RII\,ab} := \Omega \RII_{ab}$, and $f_{E\,ab} = \Omega
E_{ab} - 1/2 \RII_{ab}$ on the spectral grid.
Due to assumption~(\ref{tildek}) $f_{B\,ab}:= \Omega B_{ab}$ is
identical $0$.
\\
To calculate $\RII_{ab}$ and $E_{ab}$ on the spectral grid we apply the
procedure for dividing by $\Omega$ as outlined in the second part of
subsection~\ref{YamabeSolver}.
\\
We now have a complete set of data on the spectral grid for
$\bar\Sigma_{t_0}$.
We need to transform these data to the grid which is used in the time
evolution and which extends beyond $\bar\Sigma_{t_0}$.
Analytically we could just evaluate the sum in (\ref{SpektralSumme})
for any $(x,y,z)$ on the grid.
Numerically that is not possible:
For $|x|>1$ the absolute value of the $n$th Chebychev polynomial grows
as fast as $|x|^n$.
Although the Chebychev coefficients of analytic functions decay
exponentially fast for sufficiently large $n$, the numerically
calculated coefficients do not decay to exactly $0$ due to rounding
errors, typically not larger than $10^{-11}$, but non-vanishing.
Due to the rapid growth of $|x|^n$ for large $n$, these rounding errors
would dominate the numerical result.
The described way of extending is therefore numerically unstable.
\\
A possible solution goes as follows: For $|x|>1$ we replace the
Chebychev polynomials $T_k(x)$ by some functions $\tilde T_k(x)$ which
fuse sufficiently smooth into the Chebychev polynomials at $|x|=1$ and
which have bounded growth.
There are of course an unlimited number of options to do so, or choice
is
\begin{equation}
  \tilde T_k(x) = 
  \left\{ 
    \begin{array}{ll}
      \left(\mbox{sign}x\right)^k \cosh\left(k \, \cosh^{-1}\!\tilde{x}\right) &
        |x| > 1\\
      \cos\left(k \, \cos^{-1}\!x\right) & |x| \le 1
    \end{array}
  \right.
\end{equation}
with 
$\tilde x = \left. \left( \tan \frac{k^2}{4} \left(x-\mbox{sign}x\right)
\right) \right/ \left( \frac{k^2}{4} \right) + \mbox{sign}x$.
We call the process of calculating initial values on the time
evolution grid from the spectral grid ``extension procedure''.
\subsection{The case of spacetimes which are asymptotically Minkowski}
\hspace{-\parindent}\label{dataAsympMink}%
Asymptotically Minkowski spacetimes are spacetimes whose null infinity
has spherical cuts.
Therefore the topology of $\bar\Sigma_{t_0}$ is the one of a
three-dimensional ball (cf.\ figure~1 of I).
\\
We assume that $(x^a) = (x,y,z)$ is a coordinate system induced by $R^3$,
which we call a Cartesian coordinate system, and that the Cartesian components
$h_{\underline{a}\underline{b}}$ of the 3-metric $h_{ab}$ are smooth.
These are the coordinates in which the time evolution is done.
We call the coordinate system $(r,\vartheta,\varphi)$ defined by
\begin{subequations}
\label{PolarCoord}
\begin{eqnarray}
  x & = & r \sin\vartheta \cos\varphi \\
  y & = & r \sin\vartheta \sin\varphi \\
  z & = & r \cos\vartheta
\end{eqnarray}
\end{subequations}
the polar coordinate system $x^{a'}$.
Without loss of generality we can assume that ${\cal S}$ coincides
with the $r=1$ coordinate surface.
The interval 
$(r,\vartheta,\varphi) \in [0,1] \times [0,\pi] \times [0,2\pi[$ 
then covers the unit ball $B^3$.
\\
Using polar coordinates has the advantage, that the boundary~${\cal S}$
coincides with a coordinate surface.
On the other side we have to deal with the coordinate singularities at
$\vartheta=0$ or $\pi$ (the z-axis) and $r=0$ (the origin).
The coordinate singularities cause singular behaviour of the metric
and its derivatives and potentially lead to numerical instabilities.
To avoid these, a careful choice of the placement of the grid nodes
and special measures in the spectral transformation must be taken.
In this paper we describe what we have done. 
The reader can find a more exhaustive discussion of what can be done
in~\cite{BoM90tg}.
\\
Apart from the changes enforced by the coordinate singularities, the code
is very similar to the previous case of asymptotically A3 data.
\\[\baselineskip]
Since any interval $[a,b]$ can easily be mapped to $[-1,1]$, functions
on any interval $[a,b]$ can be represented by Chebychev polynomials.
Therefore, one could in principle work on the $r$-interval $[0,1]$.
Due to the coordinate singularities, this would not be a good idea.
Firstly, there would be a Gauss-Lobatto node directly at the origin.
And secondly, near the origin the gridpoints would be extremely dense.
Instead of representing the ball by 
$[0,1] \times [0,\pi] \times [0,2\pi[$ we use 
$[-1,1] \times [0,\pi[ \times [0,\pi[$ as minimal coordinate patch.
By the symmetries
\begin{subequations}
\label{PointIdentifications}
\begin{eqnarray}
  f(r,\vartheta+\pi,\varphi) & = & \pm f(-r,\vartheta,\varphi) \\
  f(r,\vartheta,\varphi+\pi) & = & \pm f(-r,\pi-\vartheta,\varphi),
\end{eqnarray}
\end{subequations}
where the signs depend on what tensor component is represented by $f$,
the use of the function~(\ref{CosTscheb}) instead of
(\ref{OrigTscheb}), and the identity~(\ref{CosSym}), this interval
can be mapped to the interval 
$[0,2\pi[ \times [0,2\pi[ \times [0,2\pi[$, on which everything is
$2\pi$ periodic.
Spectral transformations are performed on this extended range 
$[0,2\pi[ \times [0,2\pi[ \times [0,2\pi[$.
As in the previous section the use of the extended range is a waste of
memory, but again this is not influencing the total memory
requirements of the code.
\\
Now we have to place the gridpoints in such a way that there is no
gridpoint at the origin and on the axis.
We use $N_r+1$ gridpoints, where $N_r$ is odd, to cover the
$r$-interval $[-1,1]$:
\begin{subequations}
\begin{equation}
  \label{RGridPointsAsympMink}
  r_i = \cos(i \frac{\pi}{N_r})  \quad
  i=0,\ldots,N_r \,.
\end{equation}
Since $N_r$ is odd, there is no gridpoint at $r=0$.
The interval $[0,2\pi[$ of the extended range is then covered by 
$2 N_r$ gridpoints, i.~e.\ $4$ is not a divider of $2 N_r$, and
FFTs based on powers of $2$ cannot be used.
\\
To cover the $\vartheta$-interval $[0,\pi[$ we use $N_\vartheta$ gridpoints
which we place at
\begin{equation}
  \label{ThGridPointsAsympMink}
  \vartheta_j = j \frac{\pi}{N_\vartheta} + \frac{\pi}{2 N_\vartheta}, 
    \quad j=0,\ldots,N_\vartheta-1 \,.
\end{equation}
The shift of $\frac{\pi}{2 N_\vartheta}$ ensures, that there is no
gridpoint at the axis.
In total we have $2 N_\vartheta$ gridpoints on the $\vartheta$-interval
$[0,2\pi[$.
\\
To cover the $\varphi$-interval $[0,\pi[$ we use $N_\varphi$ gridpoints which
we place at
\begin{equation}
  \label{PhGridPointsAsympMink}
  \varphi_k = k \frac{\pi}{N_\varphi}, \quad k=0,\ldots,N_\varphi-1 \,.
\end{equation}
\end{subequations}
There is no shift needed, any value of $\varphi$ can be assumed.
In total we have $2 N_\varphi$ gridpoints on the $\varphi$-interval
$[0,2\pi[$.
\\
The discretisation of the derivative operators in ${\bf E_l}$ happens
as in~(\ref{ElDiskretisierung}), where $(x,y,z)$ is replaced by
$(r,\vartheta,\varphi)$.
\\
We now have a grid which avoids the coordinate singularities.
This is still not sufficient to avoid numerical problems.
The problems are caused by the components which behave like
powers of $1/\sin(\vartheta)$ near the axis and/or like powers of $1/r$
near the origin.
If we make a spectral transformation we always get significant
values for the highest frequencies, no matter how many basis functions
(=gridpoints) we use.
These high frequencies make the Multigrid Newton Method unstable.
To avoid the high frequencies we only apply spectral transformation to
``regularised quantities''.
What we mean by ``regularised quantities'' will become obvious from what
follows. 
\\[\baselineskip]
As free functions we give the Cartesian components
$h_{\underline{a}\underline{b}}$ of the 3-metric $h_{ab}$,
$\bar\Omega$, and $k$  as functions of the polar coordinates.
We suppose that the Cartesian components of the metric and $k$ are
given in a form which extends to the whole time evolution grid.
The polar coordinate components $h_{\underline{a}'\underline{b}'}$ of
the 3-metric $h_{ab}$ are then given by
\begin{subequations}
  \label{hPolarInCart}
  \begin{eqnarray}
    h_{rr} & = & 
      \left( 2 c_\varphi c_\vartheta h_{xz} 
        + 2 c_\vartheta h_{yz} s_\varphi \right) s_\vartheta 
        + h_{zz} (c_\vartheta)^2 
        + \left( 2 c_\varphi h_{xy} s_\varphi + h_{xx} (c_\varphi)^2
                 + h_{yy} (s_\varphi)^2 \right) (s_\vartheta)^2 \nonumber\\
           & =: & \bar h_{rr} \\
    h_{r\vartheta} & = & 
      r
      \left[ c_\varphi h_{xz} (c_\vartheta)^2 
        + h_{yz} s_\varphi (c_\vartheta)^2 
             + s_\vartheta \left( -(c_\vartheta h_{zz}) 
               + 2 c_\varphi c_\vartheta h_{xy} s_\varphi 
               + c_\vartheta h_{xx} (c_\varphi)^2 
               + c_\vartheta h_{yy} (s_\varphi)^2 \right)
           \right. \nonumber\\
       & & \left. \quad {} + \left( - c_\varphi h_{xz} 
       - h_{yz} s_\varphi\right) (s_\vartheta)^2 \right] \nonumber\\
        & =: & r \bar h_{r\vartheta} \\
    h_{r\varphi} & = &    
      r   s_\vartheta 
      \left[ \left(c_\varphi c_\vartheta h_{yz} 
          - c_\vartheta h_{xz} s_\varphi\right) 
        + \left( - c_\varphi h_{xx} s_\varphi 
          + c_\varphi h_{yy} s_\varphi + h_{xy} (c_\varphi)^2 
                 - h_{xy} (s_\varphi)^2 \right) s_\vartheta \right]  \nonumber\\
        & =: & r s_\vartheta \bar h_{r\varphi} \\
    h_{\vartheta\vartheta} & = & 
      r^2
      \left[ \left( -2 c_\varphi c_\vartheta h_{xz} 
          - 2 c_\vartheta h_{yz} s_\varphi \right) s_\vartheta 
        + 2 c_\varphi h_{xy} s_\varphi (c_\vartheta)^2 + h_{xx} (c_\varphi)^2
        (c_\vartheta)^2 
        \right.  \nonumber\\
        & & \left. \quad {}
        + h_{yy} (c_\vartheta)^2 (s_\varphi)^2 
        + h_{zz} (s_\vartheta)^2 \right]  \nonumber\\
        & =: & r^2 \bar h_{\vartheta\vartheta} \\
    h_{\vartheta\varphi} & = &
      r^2 s_\vartheta
      \left[ \left( - c_\varphi c_\vartheta h_{xx} s_\varphi 
          + c_\varphi c_\vartheta h_{yy} s_\varphi 
                    + c_\vartheta h_{xy} (c_\varphi)^2 
                    - c_\vartheta h_{xy} (s_\varphi)^2 \right) 
        + \left( - c_\varphi h_{yz} 
          + h_{xz} s_\varphi \right) s_\vartheta \right]  \nonumber\\
        & =: & r^2 s_\vartheta \bar h_{\vartheta\varphi} \\
    h_{\varphi\varphi} & = &
      r^2 s_\vartheta^2
      \left[ -2 c_\varphi h_{xy} s_\varphi + h_{yy} (c_\varphi)^2 
        + h_{xx} (s_\varphi)^2 \right]
      \nonumber\\
      & =: & r^2 (s_\vartheta)^2 \bar h_{\varphi\varphi},
  \end{eqnarray}
\end{subequations}
where $c_\vartheta := \cos\vartheta$, $c_\varphi := \cos\varphi$, $s_\vartheta :=
\sin\vartheta$, $s_\varphi := \sin\varphi$, and we have written
$h_{rr}$ although we really mean $h_{\underline{r}\underline{r}}$.
The inverse metric $h^{ab}$ can then be written as
\begin{equation}
  \label{InvPolarMetrik}
  h^{\underline{a}'\underline{b}'} =
  \left( 
    \begin{array}{ccc}
      \bar h^{rr} & 
        \frac{1}{r} \bar h^{r\vartheta} & 
        \frac{1}{r \sin\vartheta} \bar h^{r\varphi} \\
      \frac{1}{r} \bar h^{r\vartheta} & 
        \frac{1}{r^2} \bar h^{\vartheta\vartheta} & 
        \frac{1}{r^2 \sin\vartheta} \bar h^{\vartheta\varphi} \\
      \frac{1}{r \sin\vartheta} \bar h^{r\varphi} & 
        \frac{1}{r^2 \sin\vartheta} \bar h^{\vartheta\varphi} & 
        \frac{1}{(r \sin\vartheta)^2} \bar h^{\varphi\varphi} \\
    \end{array}
  \right),
\end{equation}
where $\bar h^{\underline{a}'\underline{b}'}$ is the inverse of the
matrix $\bar h_{\underline{a}'\underline{b}'}$.
\\
To calculate the polar coordinate components
$\gamma^{\underline{a}'}{}_{\underline{b}'\underline{c}'}$ of the
3-Christoffel symbols $\gamma^a{}_{bc}$ and the 3-Ricci scalar
$\RIII$, which contains derivatives of the
$\gamma^{\underline{a}'}{}_{\underline{b}'\underline{c}'}$s, 
we first calculate the first and second derivatives of the
``regularised quantities'' $\bar h_{\underline{a}'\underline{b}'}$ and
$\bar h^{\underline{a}'\underline{b}'}$ by spectral techniques, and then
put in the singular terms by hand.
This avoids the high frequency problem mentioned above.
\\
The solution $\phi$ of the Yamabe equation~(\ref{Yamabe}) is then
calculated as a function of the polar coordinates as before.
Since the scalar $\phi$ is regular on the axis and at the center, no
regularisation is needed.
\\[\baselineskip]
From the solution $\phi$ we calculate a complete set of conformal data
as follows.
The Cartesian components of $h_{ab}$ are taken from the free functions
evaluated at the gridpoints.
Then, the Cartesian components of the $\gamma^a{}_{bc}$s are obtained
by fourth order differentiation $(II/10a)$ on the time evolution
grid.
We do not use the extension procedure for the $\gamma^a{}_{bc}$s for
two reasons.
Firstly, the extension procedure is slow.
And secondly, transforming the polar coordinate components of the
$\gamma^a{}_{bc}$s to Cartesian components requires a careful
treatment of the regularity conditions.
As price for this convenience we get only 4th order approximations of
the Cartesian components of $\gamma^a{}_{bc}$ on the time evolution
grid instead of infinite order approximations.
But even if we started with infinite order data, after some time steps
everything would only be 4th order, since the time evolution scheme is
``only'' 4th order.
\\
Then we use the extrapolation procedure to calculate $\Omega$ on the
whole grid.
The spatial derivatives $\Omega_a$ of $\Omega$ are again obtained by
fourth order differentiation.
The time derivative $\Omega_0$ of $\Omega$ is calculated from the free
function on the time evolution grid directly.
\\
The scalar $\omega$ is calculated on the spectral grid and then
extended.
The quantities $\ROI_a$, $f_{\RII_{ab}}$, and $f_{E_{ab}}$ are tensor
components. 
We first calculate the polar components on the spectral grid and then
transform to Cartesian components.
To the Cartesian components of $\ROI_a$ we apply the extension
procedure.
\\
The $\Omega$ division is applied to the Cartesian components of
$f_{\RII_{ab}}$ and $f_{E_{ab}}$ on the spectral grid.
This leads to the Cartesian components of $\RII_{ab}$ and $E_{ab}$.
We complete the construction of the complete set of initial data by
extending those to the time evolution grid.
\subsection{Two examples}
\begin{figure}[htbp]
  \begin{center}
    \begin{minipage}[t]{15cm}
      \includegraphics[width=7cm]{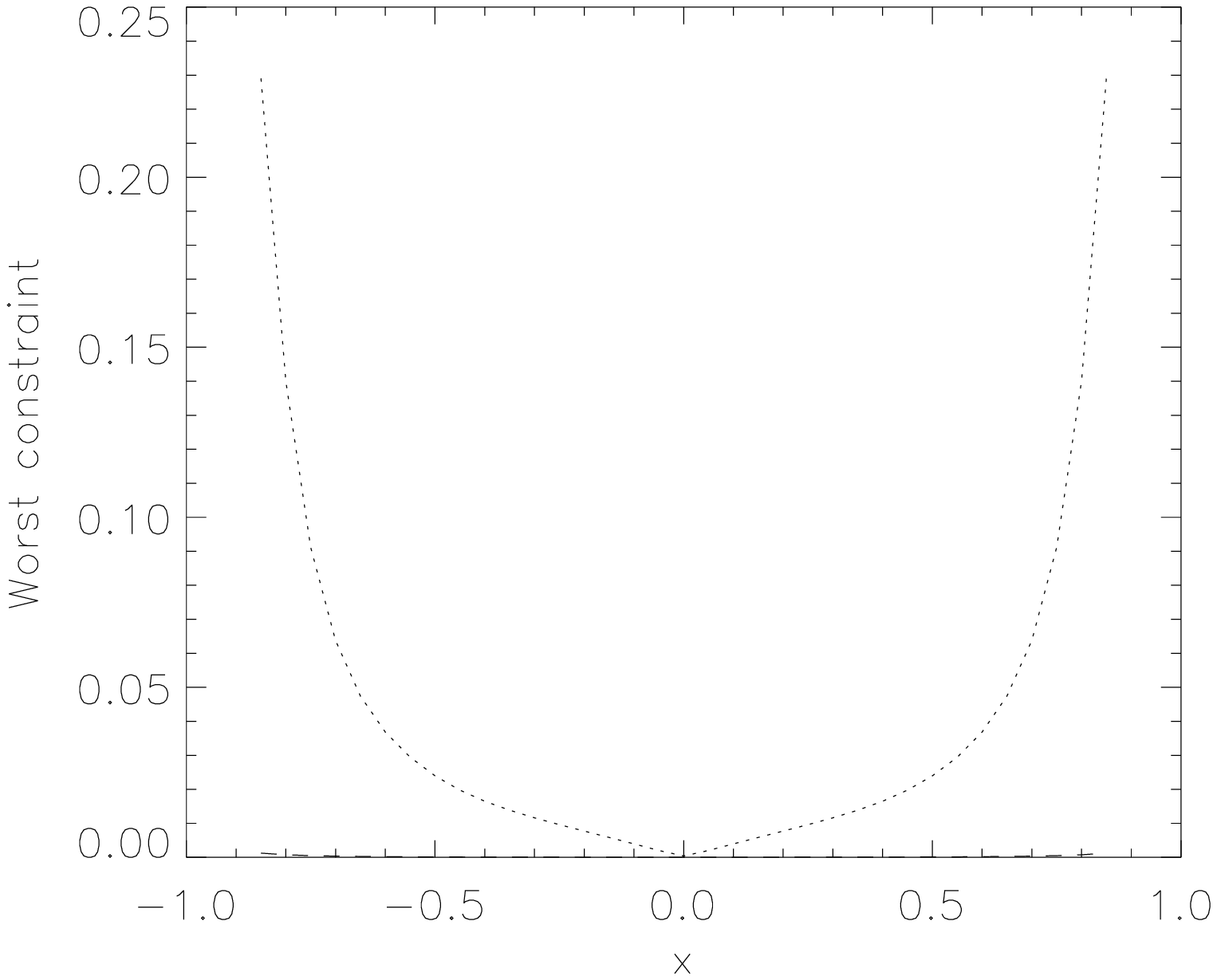}%
      \hspace{1cm}
      \includegraphics[width=7cm]{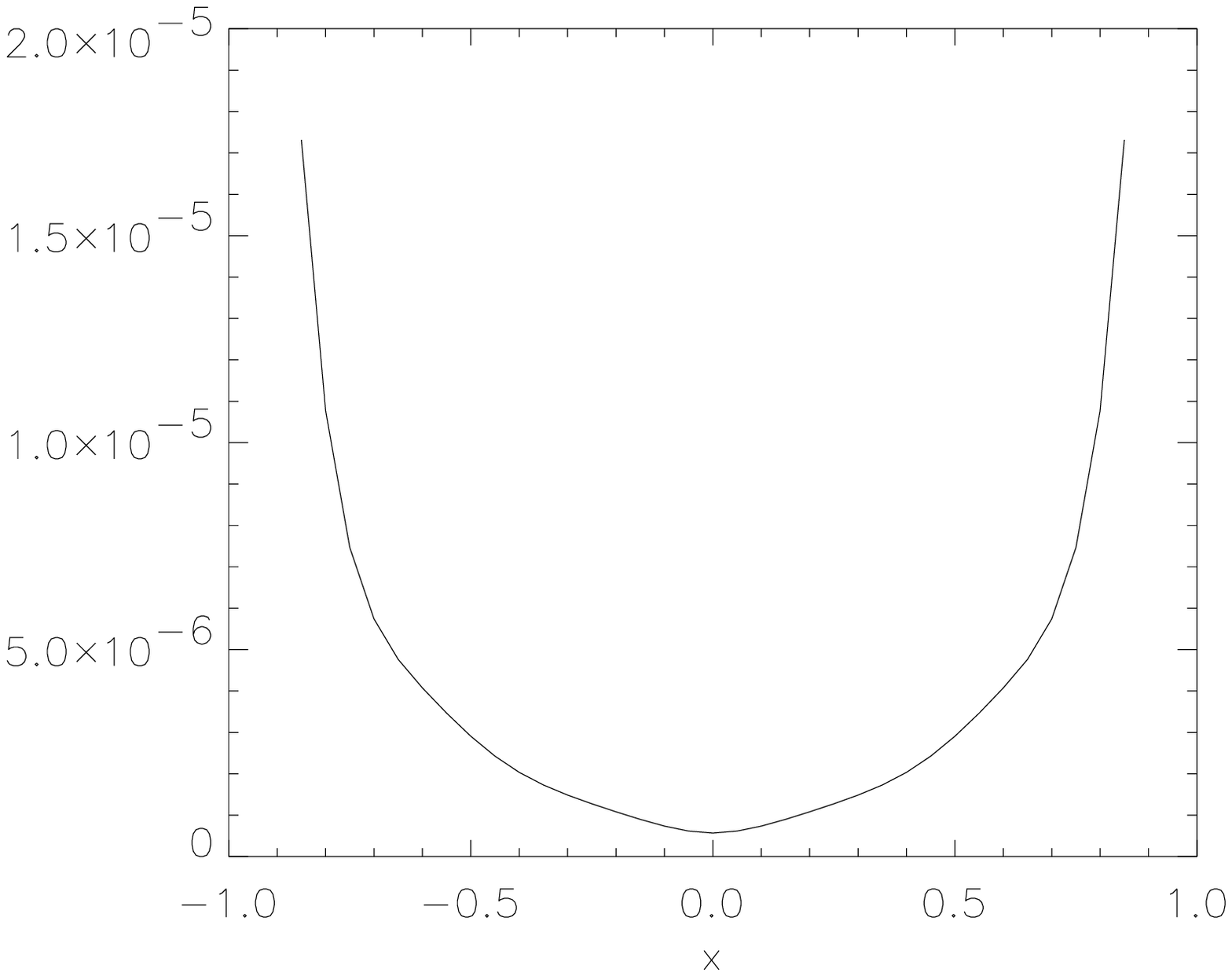}%
      \vskip0.5em
      \caption{\label{KonvergenzVerlConstr50A3}Plots of the numerical
        violation of the constraint (II/14d,$a=\underline{x}$), which
        is the constraint with the largest violation, on a $50^3$ time
        evolution grid for the data calculated from a
        $17 \times 16 \times 16$ (dotted line), a 
        $25 \times 24 \times 24$ (dashed line, almost invisible, since
        very close to the abscissa), and a 
        $33 \times 32 \times 32$ (solid line) spectral grid.}
    \end{minipage}
  \end{center}
\end{figure}

\begin{figure}[htbp]
  \begin{center}
    \begin{minipage}[t]{7cm}
      \includegraphics[width=7cm]{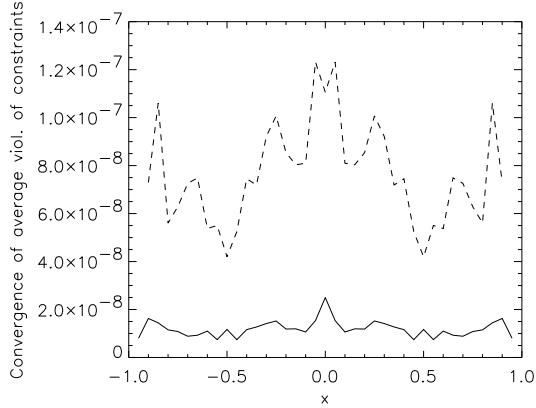}
      \vskip0.5em
      \caption{\label{KonvVerlConstrMink}Average violation of the
        constraints on a $50^3$ (dashed line) and a $100^3$ (solid line)
        time evolution grid for data calculated from a $34 \times 32
        \times 32$ spectral grid.}
    \end{minipage}
  \end{center}
\end{figure}

\hspace{-\parindent}\label{TwoExampl}%
In this subsection we give two examples for initial data calculated
with the solvers described above.
The first example is an asymptotically A3 data set, the second an
asymptotically Minkowski data set.
We have given very general free functions.
The purpose is to demonstrate the performance of the code described.
In a realistic parameter study one would probably choose much simpler
free functions.
\\[\baselineskip]
For the asymptotically A3 case we have chosen
\begin{subequations}
\begin{eqnarray}
  \bar\Omega & = &\frac{1}{2} \left( 1 - x^2 \right)
  \\
  h_{\underline{a}\underline{b}} & = &
  \left(
  \begin{array}{ccc}
    1 + \bar\Omega^2 (\sin y)^2 & 
      \frac{1}{4} \bar\Omega^2 \left( x^2 - (\sin y)^2 \right) &
      \frac{1}{4} \bar\Omega^2 (\cos z)^2  \\
    \frac{1}{4} \bar\Omega^2 \left( x^2 - (\sin y)^2 \right) &
      1 + \bar\Omega^2 (\sin z)^2 &
      \frac{1}{5} \bar\Omega^2 \cos(x\pi) \\
    \frac{1}{4} \bar\Omega^2 (\cos z)^2 &
      \frac{1}{5} \bar\Omega^2 \cos(x\pi) &
      1 + \frac{1}{2} \bar\Omega^2
  \end{array}
  \right)
  \\
  k & = & \cos y.
\end{eqnarray}
\end{subequations}
Without the $\bar\Omega^2$ terms, the choice of
$h_{\underline{a}\underline{b}}$ leads to data for an A3 spacetime.
\\
Figure~\ref{KonvergenzVerlConstr50A3} shows the convergence of the
violation of the constraint $\DIII_b E_{\underline{x}}{}^b +
\epsIII_{\underline{x}bc} k^{bd} B_d{}^c = 0$
(II/14d, with $a=\underline{x}$), which is the constraint which shows the
largest violation, for increasing spectral grid density.
For simplicity we have plotted the $L_2$ norm over $(y,z)$ as defined
in (II/30).
We observe that the violation is rapidly going to $0$ as could be
expected.
\\
As we have $90$ constraints and the violation of
(II/14d,$a=\underline{x}$) dominates for all spectral resolutions, the
average violation of constraints is more than by a factor of $50$
less.
\\
For these runs we stopped the Multigrid Newton Method iteration when
the $L_2$ residuum dropped below $10^{-7}$, 
at that time the approximation changed only by an order of $10^{-11}$
per iteration.
To achieve such a small residuum we typically need less than 20
Iterations, which means that the average improvement of the residuum
per iteration is a factor of $3$.
\\
For the $33 \times 32 \times 32$ spectral grid the average computer time
per AMG  step, the non-parallel part of the code, is about 10s for the
Yamabe equation and 5s for the $f/\Omega$ steps on our SGI Origin 2000
with MIPS R10000 processors.
\\[\baselineskip]
For the asymptotically Minkowski case we have chosen
\begin{subequations}
\begin{eqnarray}
  \bar\Omega & = &\frac{1}{2} \left( 1 - \left( x^2 + y^2 + z^2 \right) \right)
  \\
  h_{\underline{a}\underline{b}} & = &
  \left(
  \begin{array}{ccc}
    1 + \bar\Omega^2 \cos y & 
      \frac{1}{4} \bar\Omega^2 \left( (\cos x)^2 - (\sin y)^2 \right) &
      \frac{1}{4} \bar\Omega^2 (\cos z)^2  \\ 
    \frac{1}{4} \bar\Omega^2 \left( (\cos x)^2 - (\sin y)^2 \right) &
      1 + \bar\Omega^2 (\sin z)^2 &
      \frac{1}{5} \bar\Omega^2 \cos x \\
    \frac{1}{4} \bar\Omega^2 (\cos z)^2 &
      \frac{1}{5} \bar\Omega^2 \cos x &
      1 + \frac{1}{2} \bar\Omega^2
  \end{array}
  \right)
  \\
  k & = & \cos z.
\end{eqnarray}
\end{subequations}
Without the $\bar\Omega^2$ terms, the choice of
$h_{\underline{a}\underline{b}}$ leads to data for Minkowski space.
\\
Figure~\ref{KonvVerlConstrMink} shows the average violation (II/30) of
the constraints.
In this example the violation of the constraints is not dominated by a
single constraint as in the asymptotically A3 example.
The class of constraints with the largest violations are (II/14j),
then followed by (II/14d) and (II/14g).
\\
As in the asymptotically A3 example the violation of the constraints
is rapidly decreasing with the number of gridpoints in the spectral
grid.
For figure~\ref{KonvVerlConstrMink} we used a $34 \times 32 \times 32$
spectral grid.
The violation of the constraints drops by a factor of $\approx 8$, when
we refine the time evolution grid, on which we numerically evaluate
the constraints, from $50^3$ to $100^3$.
If the data were an exact solution of the constraints, this factor
would be $\approx 16$.
If the data were a bad solution of the constraints, this factor would
be $\approx 1$.
Therefore we conclude that the data must be very close to an exact
solution of the constraints.
\section{The case of multiple black hole spacetimes}
\begin{figure}[htbp]
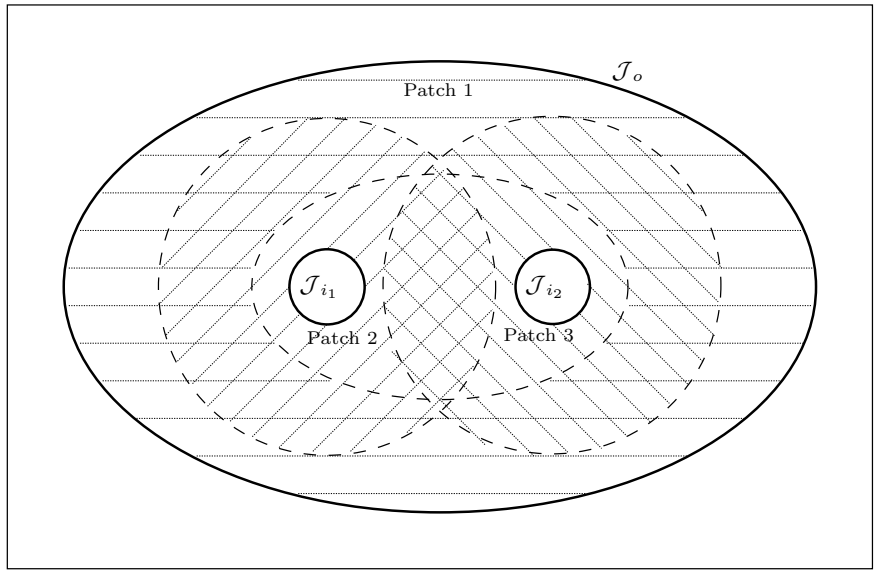

  \begin{center}
    \begin{minipage}[t]{11.6cm}
      \input nBHPatches.pstex_t
      \vskip0.5em
      \caption{\label{nBHPatches}Patches for pseudo-spectral grids to
        cover the initial slice of a two black hole spacetime. Dashed
        lines mark the interior boundaries on which boundary values
        are read off from the other patches.}
    \end{minipage}
  \end{center}
\end{figure}

\hspace{-\parindent}\label{dataAsympnBH}%
In the previous two subsections we have described in detail how initial
data for the conformal field equations can be constructed for the
asymptotically Minkowski and the asymptotically A3 case.
The cases described are spacetimes with gravitational wave content.
Depending on the ``size'' of the data, the gravitational waves
may disperse to null infinity or may interact to form black holes.
For our understanding of the Einstein equation it is certainly very
interesting to study those cases, but as models for sources of
gravitational waves they are probably not the most interesting ones.
Therefore, the method for calculating initial data as described would
only be of limited use, if it could not be extended to the case of one or
more black holes (see figures 2 and 3 of I).
In the following we describe how one can build a code to calculate
data describing multiple black holes.
\\[\baselineskip]
The case of one black hole with radiation (asymptotically
Schwarzschild) is a straightforward extension of the asymptotically
Minkowski case, since we can also use polar coordinates.
Without loss of generality we can choose $\bar\Omega$ such that the
inner ${\cal S}_i$ coincides with the coordinate sphere at $r_i>0$ and the
outer ${\cal S}_o$ coincides with the coordinate sphere at $r_o>r_i$.
Instead of using $[-1,1] \times [0,\pi[ \times [0,\pi[$ to cover the
physical part of the grid, we use 
$[r_i,r_o] \times [0,\pi[ \times [0,2 \pi[$.
Regularisation is even easier, since $r=0$ is not part of
$\bar\Sigma_{t_0}$, there is only an axis singularity but not an origin
singularity to deal with.
\\
When extending to the time evolution grid inside ${\cal S}_i$ more
care is needed than in the asymptotically Minkowski case, since we
extend onto the $r=0$ coordinate singularity of the time evolution
grid.
When extending towards $r=0$ the limit must not depend on
the direction.
\\[\baselineskip]
The case of two black holes is significantly more complex.
Since we do not see how to cover $\bar\Sigma_{t_0}$ with a coordinate
system for which we can also provide a spectral basis and fast
spectral transformation algorithms, we suggest to use domain
decomposition techniques.
\\
In figure~\ref{nBHPatches} we give our three coordinate patches.
One boundary of each patch coincides with a part of ${\cal S}$, the
other, the dashed lines, in future called ${\cal G}$, lies inside
$\bar\Sigma_{t_0}$ and inside at least one of the other patches.
If we knew boundary values on ${\cal G}$, we would have three
copies of the asymptotically Schwarzschild case --- the elliptical shape
of patch~1 is easily dealt with by a straightforward coordinate
transformation.
In the case of the $f/\Omega$ divider we do know the boundary values
at ${\cal G}$, namely $f/\Omega$, for the Yamabe equation we do not
the boundary values at ${\cal G}$.
\\
To provide boundary values on ${\cal G}$ we can use the Schwarz
alternating procedure as described in section~6.4.1 of~\cite{QuV97NA}:
When solving the equation in one patch we read off the boundary values
on ${\cal G}$ from the grids covering the other patches.
By iteratively solving the equation on all patches we calculate a
solution on $\bar\Sigma_{t_0}$.
The book~\cite{QuV97NA} contains a proof, that the Schwarz alternating
procedure converges for the Laplace equation.
The convergence rate depends on the overlap of the patches,
the larger the overlap the better.
In the overlap regions of our patches equation~(\ref{Yamabe}) is a
normal elliptic equation, the principal part does not degenerate
there.
Therefore, it should in principle be possible to also prove
convergence of the Schwarz alternating procedure for
equation~(\ref{Yamabe}).
%
%
%

%% file: nBHPatches.pstex_t
\begin{picture}(0,0)%
\epsfig{file=nBHPatches.pstex}%
\end{picture}%
\setlength{\unitlength}{2072sp}%
\begingroup\makeatletter\ifx\SetFigFont\undefined%
\gdef\SetFigFont#1#2#3#4#5{%
  \reset@font\fontsize{#1}{#2pt}%
  \fontfamily{#3}\fontseries{#4}\fontshape{#5}%
  \selectfont}%
\fi\endgroup%
\begin{picture}(10394,6794)(204,-6608)
\put(5401,-916){\makebox(0,0)[b]{\smash{\SetFigFont{7}{8.4}{\familydefault}{\mddefault}{\updefault}Patch 1}}}
\put(3826,-3886){\makebox(0,0)[lb]{\smash{\SetFigFont{7}{8.4}{\familydefault}{\mddefault}{\updefault}Patch 2}}}
\put(7021,-3841){\makebox(0,0)[rb]{\smash{\SetFigFont{7}{8.4}{\familydefault}{\mddefault}{\updefault}Patch 3}}}
\put(7471,-691){\makebox(0,0)[lb]{\smash{\SetFigFont{9}{10.8}{\familydefault}{\mddefault}{\updefault}$\Scri_o$}}}
\put(3736,-3256){\makebox(0,0)[lb]{\smash{\SetFigFont{9}{10.8}{\familydefault}{\mddefault}{\updefault}$\Scri_{i_1}$}}}
\put(6436,-3256){\makebox(0,0)[lb]{\smash{\SetFigFont{9}{10.8}{\familydefault}{\mddefault}{\updefault}$\Scri_{i_2}$}}}
\end{picture}

%% file: Zusammenfassung.tex
%
%
\section{Conclusion}
\hspace{-\parindent}%
In this paper we have described a highly efficient and accurate scheme
to calculate data for the conformal field equations without making any
symmetry assumptions.
The data are specified by giving a boundary defining function
$\bar\Omega$ and the six components of the conformal 3-metric
$h_{ab}$.
\section*{Acknowledgement}
\hskip-\parindent{}%
I would like to thank H.~Friedrich for his help and support and
J.~Frauendiener for many helpful discussions on pseudo-spectral methods.
\\
B.~Schmidt, M.~Weaver, and S.~Husa should also be mentioned for many
discussions.
\\
I also acknowledge K.~St\"uben from the Gesellschaft f\"ur
Mathematik und Datenverarbeitung, who put the Algebraic Multigrid
Library AMG at my disposal, and M.~Frigo and S.~G.~Johnson who wrote FFTW
and made it publically available for the scientific community.
%
%
%
%

%% file: biblio.tex
\bibliography{biblio}
\bibliographystyle{prsty}